# A Simulation Modeling Approach for Optimization of Storage Space Allocation in Container Terminal

Gamal Abd El-Nasser A. Said, El-Sayed M. El-Horbaty

*Abstract*—Container handling problems at container terminals are NP-hard problems. This paper presents an approach using discrete-event simulation modeling to optimize solution for storage space allocation problem, taking into account all various interrelated container terminal handling activities. The proposed approach is applied on a real case study data of container terminal at Alexandria port. The computational results show the effectiveness of the proposed model for optimization of storage space allocation in container terminal where 54% reduction in containers handling time in port is achieved.

## I. INTRODUCTION

THE increasing development of containerized transportation and the global competition among different ports are the main causes for the wide attention given by researchers to seaport container terminals. As a result of increasing competitions between container ports, improving efficiency in container terminals has become an important and immediate challenge for all managers in order to gain higher competitiveness. The optimization of maritime container terminals is a research issue that has received, in recent years, increasing attention by the international research community. A container terminal is a transit point for containerized goods between sea vessels and land transportation modes, such as trucks [3], [22], [24]. One of the most important performance measures in container terminals is the operation time. In most container terminals, there are three main types of handling equipment involved in the loading/discharging process: Quay Cranes (QCs), Yard Trucks (YTs), and Yard Cranes (YCs). QCs are in charge of quayside operation; YCs focus on operation in yard; YTs are in charge of transportation between quayside and yard. Therefore, the integrated of QC–YT–YC is of great impact upon cost reduction and efficiency improvement of the whole port [7], [8], [15], [30]. After arrival at the port, a container vessel is assigned to a berth equipped with cranes to load and unload containers. Unloaded import containers are transported to yard positions near to the place where they will be transshipped next. The layout of the container yard is an important factor in the efficiency of storage in a container terminal, in which loading and unloading of containers on the stack is performed by yard cranes or transfer cranes. Thus, designing the layout of container yards where transfer cranes and yard trucks are one method used for container stacking is a problem [1], [2], [31].

A container terminal is a zone of the port where sea-freight dock on a berth and containers are loaded, unloaded and stored in a buffer area called yard. Inbound containers are unloaded from container ships by quay cranes and then transported by internal trucks to storage yard where they are stacked by yard cranes to their allocated positions waiting for the consignees to pick up. Outbound containers are handled in the opposite direction. Shippers send the containers into the terminal and yard cranes store them in their allocated positions. Later they are retrieved by yard cranes and transported by trucks to quayside where they are loaded onto ships by quay cranes [11]. The whole terminal operation is very complex and involves different types of equipments. A terminal can therefore be ideally divided into two areas, the quayside and the yard. The quayside is made up of berths for vessels and quay cranes (QC) which move containers. The yard serves as a buffer for loading, unloading and transshipping containers and it is typically divided into blocks: each container block is served by one or more yard cranes (YC) [12].

The rest of this paper is organized as follows: A literature review of the methodologies for solving the storage space allocation problem are provided in section II. Section III discusses the methodology developed using discrete-event simulation modeling to optimize solution for storage space allocation problem. Computational results are given in Section IV. Finally, in Section V the conclusion and future work are given.

## II. LITERATURE REVIEW

In view of the increasing importance of marine transportation systems, issues related to container terminal operations are getting more and more attention. In the last decades, many approaches have been developed to solve container terminals problems. The need for optimization using methods of operations research in container terminal operation has become more and more important in recent years. This is because the logistics especially of large container terminals has already reached a degree of complexity that further improvements require scientific methods; Stacking logistics has become a field of increasing importance because more and more containers have to be stored in ports as container traffic grows continuously and space is becoming a scarce resource [10], [19]. The efficiency of storage and stacking containers is one of the most important factors for a container terminal [13]. In container terminal operations, the storage yard plays an

Gamal Abd El-Nasser A. Said is with the Computer science Department, Faculty of Computer & Information Sciences, Ain Shams University, Egypt (e-mail: jamalabdelnasser@hotmail.com).
El-Sayed M. El-Horbaty is Head of Computer science Department, Faculty of Computer & Information Sciences, Ain Shams University, Egypt.

important role for a terminal's overall performance because it links the seaside and landside and serves as the buffer area for storing containers. Therefore, storage and stacking logistics has become a field that increasingly attracts attentions in both academic and practical research during the recent years [25].

Modeling and simulation are essential tools for the design and analysis of container terminals. A computer model can emulate the activities at various levels of details and capture the essential interactions among the subsystems. Analysis based on the simulation is particularly useful for designing new terminals, making modifications to existing terminals, and evaluating the benefits of new resources or impacts of operation policies [21]. Discrete Event Simulation (DES) models help to achieve several aims: overcome mathematical limitations of optimization approaches, support computer-generated strategies/policies and make them more understandable, and support decision makers in daily decision processes through a ''what if'' approach [28].

Legat et al. present queuing-based representation of the housekeeping process in a real container terminal and solve it by discrete-event simulation to i) assess the efficiency of the housekeeping operations under unforeseen events or process disturbances and ii) estimate the related productivity and waiting phenomena which, in turn, affect the vessel turn-around time. Ports with multimodal transportation systems are in particular complex as they typically operate with ships arriving to one or more terminals, multiple quay cranes, rubber tyred gantry cranes, and trucks delivering containers of different types to terminals. With several resources of different types working and interacting, the system can be so complex that it is not easy to predict the behavior of the system and its performance metrics without the use of simulation [17]. Abd El-Nasser et al. presented a simulation model to study berth allocation problem. The proposed approach is applied on a real data from Container Terminal at El-Dekheilla port. Computational experiments were conducted to analyze the performance of container terminal operation. The results show that the proposed approach reduced the ship turnaround time in El-Dekheilla port [6].

Container handling problems at container terminals are NP-Hard, stochastic, nonlinear and combinatorial optimization problems [27], [29]. If the search space is large, it will become difficult to solve the optimization problem by using conventional mathematics or using numerical induction techniques. For this reason, many meta-heuristic optimization methods have been developed to solve such difficult optimization problems. The meta-heuristic approaches are not guaranteed to find the optimal solution since they evaluate only a subset of the feasible solutions, but they try to explore different areas in the search space in a smart way to get a near-optimal solution in less cost and time [16]. Met-heuristics algorithms have been used to solve optimization problems, Among all of the heuristic algorithms such as: genetic Algorithm, tabu Search, and simulated annealing, genetic algorithms (GAs) are in wide application because of their ability to locate the optimal solution in the global solution space [4], [9], [23]. Abd El-Nasser et al. presents a comparative study between Meta-heuristic algorithms: Genetic Algorithm, Tabu Search, and Simulated Annealing for solving Quadratic Assignment Problem. The computational results show that genetic algorithm has a better solution quality than the other Meta-heuristic algorithms for solving Quadratic Assignment Problems [18].

Han et al. formulated a mixed-integer programming (MIP) model to determine the number of incoming containers, appropriate storage locations and the smallest number of yard cranes needed for deployment in each shift. The authors developed a tool able to provide a holistic and systematic approach to the problem. A model developed in the study based on a MIP formulation was particularly useful for transshipment hubs. An iterative improvement method based on the sequential method for the yard allocation problem was also provided to solve the MIP model. However, the method cannot obtain optimal solutions in every situation [32]. Mohammad. B. et al. solved an extended Storage Space Allocation Problem (SSAP) in a container terminal by GA. The objective of the SSAP developed is to minimize the time of storage and retrieval time of containers [14]. Ayachi et al. present a genetic algorithm (GA) to solve the container storage problem in the port. The problem is studied with different container types. The objective aims to determine an optimal containers arrangement, which respects customers' delivery deadlines, reduces the re-handle operations of containers and minimizes the stop time of the container ship. An adaptation of the genetic algorithm to the container storage problem is detailed and some experimental results are presented and discussed [26]. A simulation optimization model for scheduling loading operations in container terminals is developed to find good container loading sequences which are improved by a genetic algorithm through an evaluation process by simulation model to evaluate objective function of a given scheduling scheme [5]. Sriphrabu et al. propose a developed simulation model for stacking containers in a container terminal through developing and applying a genetic algorithm (GA) for containers location assignment with minimized total lifting time and increased service efficiency of the container terminals. The application obtained from the genetic algorithm shows the appropriate containers location assignment based on the arrival of containers at a container terminal and correlates it with the order of the containers loaded onto container ships [20].

A container yard serves as a buffer area for the loading, unloading, and transshipping of containers, and is typically divided into blocks. Assigning storage space to import and export containers at the yard of a terminal, which is refereed as the yard allocation problem. This problem consists of assigning block spaces at the yard to containers arriving to the terminal, with the aim of obtaining a better space utilization of the yard, minimizing vessels handling time and maximizing the throughput of the terminal. Many approaches have been developed to solve the storage space allocation problem separately without regard to various interrelated container terminal handling activities. This paper proposes a developed an approach using discrete-event simulation modeling to optimize solution for storage space allocation problem, taking into account all various interrelated container terminal handling activities. The proposed approach is applied on a real case study data of container terminal at Alexandria port.

III. METHODOLOGY

Literature review has shown that various modeling are proposed to describe the operations carried out in container terminals and several literature contributions developed models to solve storage space allocation problem. Although container terminal problems are related to each other, many researches tackled storage space allocation problem separately. The layout of container terminal is presented in Fig. 1. Discrete-event systems are well suited to representing various activities performed in container terminals.

In this study, we have built a simulation model that can be used to simulate container terminal operation. The objective of the model is to optimize solution for storage space allocation problem, taking into account all various interrelated container terminal handling activities. The proposed approach is applied on a real case study data of container terminal at Alexandria port.

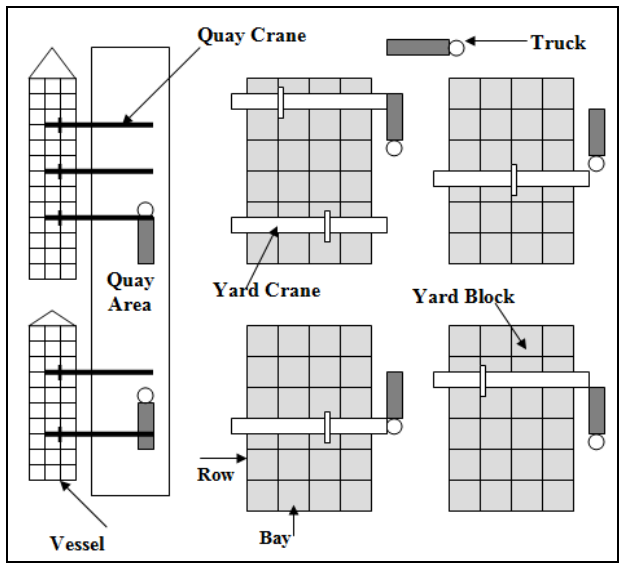

Fig. 1 Container Terminal layout

A. Case Study: Container Terminal at Alexandria Port

Egypt is a country uniquely located at the crossroads of the three continents with 2000 km. of coastal frontiers. Its north border is located on the Mediterranean Sea, where one of the oldest ports, the port of Alexandria. Alexandria Container Handling Company (ACHC) is the oldest container handling company in Egypt. Alexandria container terminal consists of 530 meters of quay length with a depth of 14 meters, terminal area is 163000 m2. The terminal equipments are 5 quay cranes, 8 RTG, 15 Top Lift Trucks, 8 Empty Handler Side Spreader, and 25 trucks [34].

B. Data Collection and Preparation

The data used in this study was obtained from Alexandria container terminal. The yards name, type of storage yards, and yards storage capacity at Alexandria container terminal are summarized in Table I. The total capacity of storage yards = 15666 (TEUs) (twenty-foot equivalent unit).

TABLE I
STORAGE YARDS CAPACITY OF CONTAINER TERMINAL

| Storage yard type | Yard Name | Storage capacity (TEU) | Yard Capacity (TEU) |
|---|---|---|---|
| Export | Export | 1050 | 2580 |
| | Export A | 600 | |
| | Export B | 930 | |
| Import | Import B | 930 | 8980 |
| | C | 1860 | |
| | D | 1920 | |
| | H | 1950 | |
| | F | 1536 | |
| | Yard 65 | 448 | |
| | Yard 32 | 336 | |
| Hazardous | Hazardous | 486 | 486 |
| Reefers | Reefers | 500 | 500 |
| Empty | Empty | 540 | 3120 |
| | Yard 22 | 2400 | |
| | Yard 60 | 180 | |
| Total Terminal Capacity (TEU) | | | 15666 |

The daily "log sheets" of the operation and planning departments at Alexandria container terminal include the detailed information for each ship arrived to the port; ship name, ship length, bay plan, number of containers, operation type (discharge/load), size of containers, containers types, operation time (start/end), date and time of arrival, berthing, and departure from port. Actual data for one week operation from 3/3/2014 to 9/3/2014 which required for our proposed model were selected (see Table II). A simulation model was developed and verified and validated using the terminal operational parameters.

C. Simulation Tool

In this study, the problems being studied are stochastic, dynamic and discrete, so this discrete-event simulator is suitable. The discrete-event simulation tool – Flexsim (Flexsim Software Products Inc) is employed for implementing the simulation model. The motivation for adopting a simulation tool for the study is as follows. Firstly, systems can be modeled as discrete with Flexsim. Another benefit it offers is its high degree of openness and flexibility. Secondly, a distinct advantage of Flexsim software over other software is that it comes with flexsim CT, a library specifically designed for simulating container terminal operation [33]. A snapshot of container Terminal from the simulation model run is shown in Fig. 2.

IV. COMPUTATIONAL RESULTS

The input data for simulation models are based on a real case study data of container terminal at Alexandria port for one week operation from 3/3/2014 to 9/3/2014. This involved

12 ship arrivals, with number of containers (discharge/load) =9531 TEU with different sizes (20ft and 40ft) and different types (full, hazardous, reefers, and empty) (see Table II).

Implementation of the model was run on a Laptop with the following configurations: i3 CPU 2.4 GHZ, 4.0 GB RAM, Windows 7.

TABLE II
CONTAINER TERMINAL ACTUAL OPERATION

| Ship No. | operation date (start) | operation date (end) | ship length (m) | Operation (Discharge / Load) | Full 20 | Full 40 | Hazardous 20 | Hazardous 40 | Reefers 20 | Reefers 40 | Empty 20 | Empty 40 | Total 20 | Total 40 | Total containers (discharge/load) TEU | Actual operation time (minutes) | operation time Our Proposed Model (minutes) |
|---|---|---|---|---|---|---|---|---|---|---|---|---|---|---|---|---|---|
| 1 | 03/03/14 04:00AM | 03/03/14 08:00 M | 186 | Discharge | 39 | 140 | 1 | | | | | | 40 | 140 | 320 | 960 | 306 |
| | | | | Load | 39 | 42 | | | | | | 1 | 39 | 43 | 125 | | |
| 2 | 04/03/14 01:00AM | 04/03/14 08:00AM | 183 | Discharge | 59 | 184 | 1 | 8 | 1 | 17 | | | 61 | 209 | 479 | 420 | 384 |
| | | | | Load | 6 | | | | | | 40 | 110 | 46 | 110 | 266 | | |
| 3 | 04/03/14 09:30PM | 05/03/14 07:30AM | 187 | Discharge | 81 | 57 | 21 | 2 | | 4 | | 131 | 102 | 194 | 490 | 600 | 312 |
| | | | | Load | 90 | 17 | | | | | 2 | 20 | 92 | 37 | 166 | | |
| 4 | 05/03/14 02:15PM | 05/03/14 11:59PM | 166 | Discharge | 20 | 7 | 1 | | | | | 78 | 21 | 85 | 191 | 584 | 269 |
| | | | | Load | 4 | 82 | | | | | 101 | 77 | 105 | 159 | 423 | | |
| 5 | 06/03/14 10:30AM | 07/03/14 09:00AM | 98 | Discharge | 120 | 74 | 14 | 2 | | | | | 134 | 76 | 286 | 1350 | 460 |
| | | | | Load | 2 | | | | | | 45 | 149 | 47 | 149 | 345 | | |
| 6 | 06/03/14 02:15PM | 07/03/14 09:00AM | 136 | Discharge | 120 | 77 | 19 | 10 | | | | 101 | 139 | 188 | 515 | 1125 | 645 |
| | | | | Load | 45 | 101 | | | | 1 | 4 | | 49 | 102 | 253 | | |
| 7 | 07/03/14 02:30PM | 07/03/14 11:59PM | 101 | Discharge | 54 | 61 | 8 | | | | | | 62 | 61 | 184 | 569 | 237 |
| | | | | Load | 54 | 1 | | | | | | 53 | 54 | 54 | 162 | | |
| 8 | 07/03/14 07:45PM | 08/03/14 04:00PM | 157 | Discharge | 267 | 178 | 9 | 9 | 14 | 46 | | | 290 | 233 | 756 | 1215 | 474 |
| | | | | Load | | | | | | | 12 | 158 | 12 | 158 | 328 | | |
| 9 | 08/03/14 02:30PM | 09/03/14 04:00AM | 223 | Discharge | 108 | 165 | 18 | 6 | 1 | 74 | | | 127 | 245 | 617 | 810 | 307 |
| | | | | Load | 16 | 19 | | | | | 40 | 7 | 56 | 26 | 108 | | |
| 10 | 09/03/14 01:00AM | 09/03/14 03:30AM | 128 | Discharge | 3 | 11 | | | | 2 | | 30 | 3 | 43 | 89 | 150 | 95 |
| | | | | Load | 10 | 3 | | | | | | | 10 | 3 | 16 | | |
| 11 | 09/03/14 10:15AM | 10/03/14 02:00AM | 194 | Discharge | 185 | 239 | 23 | 3 | | | 4 | | 212 | 242 | 696 | 945 | 723 |
| | | | | Load | 180 | 200 | 8 | | | 2 | 125 | 103 | 313 | 305 | 923 | | |
| 12 | 09/03/14 03:15PM | 11/03/14 11:00AM | 186 | Discharge | 111 | 198 | 6 | 1 | | | | 400 | 117 | 599 | 1315 | 2625 | 2191 |
| | | | | Load | 48 | 27 | | | | 89 | 40 | 79 | 88 | 195 | 478 | | |
| | | | | | | | | | | | | | 2239 | 3696 | 9531 | 11353 | 6403 |

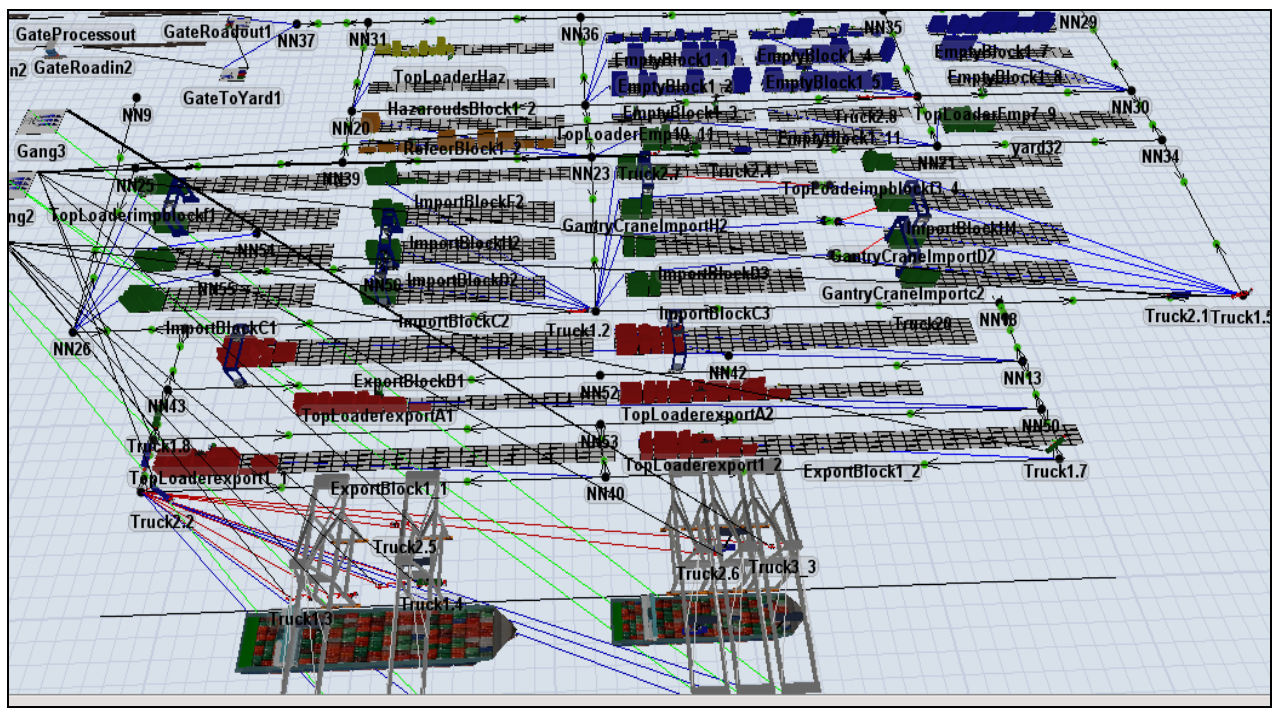

Fig. 2 Snapshot of container Terminal from the simulation model run

## A. Experimental Setting

- All resources from the same type have the same specifications.
- A vessel is allocated to the berth depending on First Come First Serve rule.

Table II shows the results of our proposed simulation model on a real case study data of container terminal at Alexandria port operation for one week from 3/3/2014 to 9/3/2014. The results shows that our proposed simulation model achieved a total containers handling time = 6403 minutes, where the actual containers handling time of the collected data=11353 minutes (Table II). The computational results show the effectiveness of the proposed model for optimization of storage space allocation in container terminal where 54% reduction in containers handling time in port is achieved. Fig. 3 illustrates a graph comparison between both actual and proposed model containers handling time versus number of (discharge/load) containers (TEU). From Fig. 3 it is obvious that our proposed simulation model enhanced the handling time and achieved better results than the methods that container terminal use.

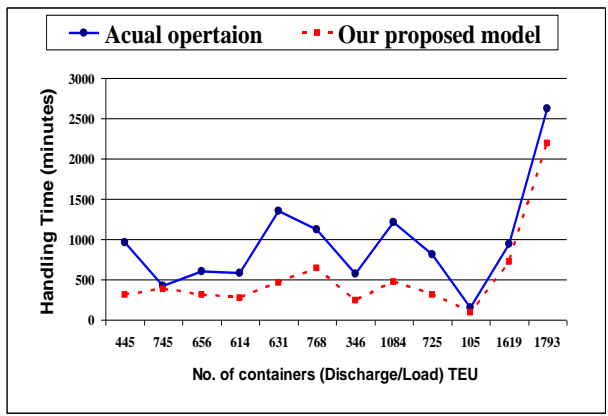

Fig. 3 Comparison between actual and proposed model containers handling time

## V. CONCLUSION AND FUTURE WORK

In this paper, we proposed a simulation model that can be used to optimize storage space allocation in container terminal using simulation methodology, taking into account all various interrelated container terminal handling activities. The proposed approach is applied on a real case study data of container terminal at Alexandria port. Computational results show the effectiveness of the proposed model for optimization of storage space allocation in container terminal where 54% reduction in containers handling time in port is achieved.

In the future, analytical and empirical evaluations will be done for optimization of container handling activities as a whole using simulation based optimization technique which combines optimization technique and discrete-event simulation modeling.

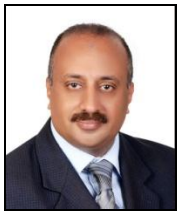
**Gamal Abd El-Nasser A. Said**: He received his M.Sc. (2012) in computer science from College of Computing & Information Technology, Arab Academy for Science and Technology and Maritime Transport (AASTMT), Egypt and B.Sc (1990) from Faculty of Electronic Engineering, Menofia University, Egypt. His work experience as a Researcher, Maritime Researches & Consultancies Center, Egypt. Computer Teacher, College of Technology, Kingdom Of Saudi Arabia and Lecturer, Port Training Institute, (AASTMT), Egypt. Now he is Ph.D. student in computer science, Ain Shams University. His research areas include optimization, discrete-event simulation, and artificial intelligence.

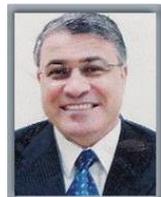
**Professor El-Sayed M. El-Horbaty**: He received his Ph.D. in Computer science from London University, U.K., his M.Sc. (1978) and B.Sc (1974) in Mathematics From Ain Shams University, Egypt. His work experience includes 39 years as an in Egypt (Ain Shams University), Qatar, Qatar University) and Emirates (Emirates University, Ajman University and ADU University). He Worked as Deputy Dean of the faculty of IT, Ajman University (2002-2008). He is working as a Vice Dean of the faculty of Computer & Information Sciences, Ain Shams University (2010-Now). Prof. El-Horbaty is current areas of research are parallel algorithms, combinatorial optimization, image processing. His work appeared in journals such as Parallel Computing, International journal of Computers and Applications (IJCA), Applied Mathematics and Computation, and International Review on Computers and software.